\makeatletter
\@ifundefined{@parse@version@dash}{%
\def\@parse@version#1{\@parse@version@0#1}
\def\@parse@version@#1/#2/#3#4#5\@nil{%
\@parse@version@dash#1-#2-#3#4\@nil}
\def\@parse@version@dash#1-#2-#3#4#5\@nil{%
  \if\relax#2\relax\else#1\fi#2#3#4 }
}{}
\makeatother

\documentclass[%
 reprint,
 amsmath,amssymb,
 aps,
onecolumn
]{revtex4-2}

\usepackage[colorlinks]{hyperref}

\usepackage{graphicx}
\usepackage{dcolumn}
\usepackage{bm}

\usepackage{color}

\hypersetup{
	setpagesize=false,
	bookmarksnumbered=true,%
	bookmarksopen=true,%
	colorlinks=true,
	linkcolor=blue,
	citecolor=blue,
	urlcolor=black,
}

\begin{document}

\title{
Pressure Evolution of Magnetic Structure and Quasiparticle Excitations \\ in Anisotropic Frustrated Zigzag Chains
}%

\author{Fumiya Hori$^1$}
\email{hori.fumiya.36s@st.kyoto-u.ac.jp}
\author{Hiroyasu Matsudaira$^1$}
\author{Shunsaku Kitagawa$^1$}
\author{Kenji Ishida$^1$}
\email{kishida@scphys.kyoto-u.ac.jp}
\affiliation{$^1$Department of Physics, Kyoto University, Kyoto 606-8502, Japan}

\author{Hiroto Suzuki$^2$}
\author{Takahiro Onimaru$^2$}
\affiliation{$^2$\mbox{Department of Quantum Matter, Graduate School of Advanced Science and Engineering, Hiroshima University,} \\ Higashihiroshima 739-8530, Japan}

\begin{abstract}
\section*{Abstract}
{
Frustrated magnetic systems with anisotropic exchange interactions have been recognized as key platforms for discovering exotic quantum states and quasiparticles.
In this study, we report the pressure evolution of magnetic structures and quasiparticle excitations in the frustrated semiconductor YbCuS$_2$, characterized by Yb$^{3+}$ zigzag chains with competing exchange interactions.
At ambient pressure, YbCuS$_2$ exhibits a magnetic transition at $T_{\rm N} \sim 0.95$~K, forming an incommensurate helical magnetic order. Under hydrostatic pressure of 1.6~GPa, $T_{\rm N}$ increases to 1.17 K, and the magnetic structure changes to a commensurate one, which can be regarded as an odd-parity magnetic multipole order.
Remarkably, pressure suppresses the gapless quasiparticle excitations. These findings suggest that pressure alters the exchange interactions between the Yb ions, affecting both the magnetic ground state and the quasiparticle excitations.
Our results highlight the pivotal role of anisotropic interactions in one-dimensionality to stabilize the complex quantum phases, offering insights into the interplay among frustration, dimensionality, multipoles and emergent quasiparticles.
}

\end{abstract}

\maketitle

\section{Introduction}
Frustrated magnets with anisotropic exchange interactions have garnered significant attention in recent years~\cite{Yb_frustration, 4d_5d_frustration, Kitaev, Kitaev-review, Kitaev-solid, Yb_triangular_1, Yb_triangular_2, Yb_triangular_3, Pyrochlore_photon_1, Pyrochlore_photon_2, Pyrochlore_photon_3, Saito1, Saito2, Saito3}.
This is because such frustrated systems exhibit non-trivial ground states and exotic quasiparticle excitations which are not anticipated in simple $S = 1/2$ Heisenberg models with isotropic exchange interactions.
For example, the anisotropic Kitaev interaction on a honeycomb lattice gives rise to a quantum spin liquid as a ground state~\cite{Kitaev, Kitaev-review, Kitaev-solid}.
In the Kitaev quantum spin liquid, the elementary excitations, including Majorana fermions, visons, and anyons, are anticipated~\cite{Kitaev, Kitaev-review, Kitaev-solid}.
Similarly, even though the ground state of the $S = 1/2$ Heisenberg triangular antiferromagnets is theoretically expected to be the 120$^{\circ}$ Neel ordered state~\cite{TLHAF1, TLHAF2, TLHAF3}, highly anisotropic and bond-dependent exchange interactions may stabilize the quantum spin liquid state; spinon Fermi-surface states or $Z_2$ spin liquid~\cite{Yb_triangular_1, Yb_triangular_2, Yb_triangular_3}.
The ground state for $S = 1/2$ pyrochlore lattice antiferromagnets with XXZ anisotropy can be effectively described by quantum electrodynamics, which hosts three different quasiparicles; magnetic and electric monopoles, and photons~\cite{Pyrochlore_photon_1, Pyrochlore_photon_2, Pyrochlore_photon_3}.
These fascinating phenomena have been studied experimentally in 4$f$-, 4$d$-, and 5$d$-electron insulating/semiconducting systems with strong spin-orbit coupling, characterized by the highly anisotropic interactions~\cite{YbMgGaO4-1, YbMgGaO4-2, YbMgGaO4-3, NaYbSe2, NaYbSe2_spinon_fermi_surface, Yb2Ti2O7_thermal_conductivity, RuCl3-kasahara, RuCl3-kasahara-2, RuBr3, Ohmagari1, Ohmagari2, Hori2022, Hori2023, RAgSe2, Hori_YbAgSe2}.

\begin{figure}[t]
\centering
\includegraphics[width=8.5cm]{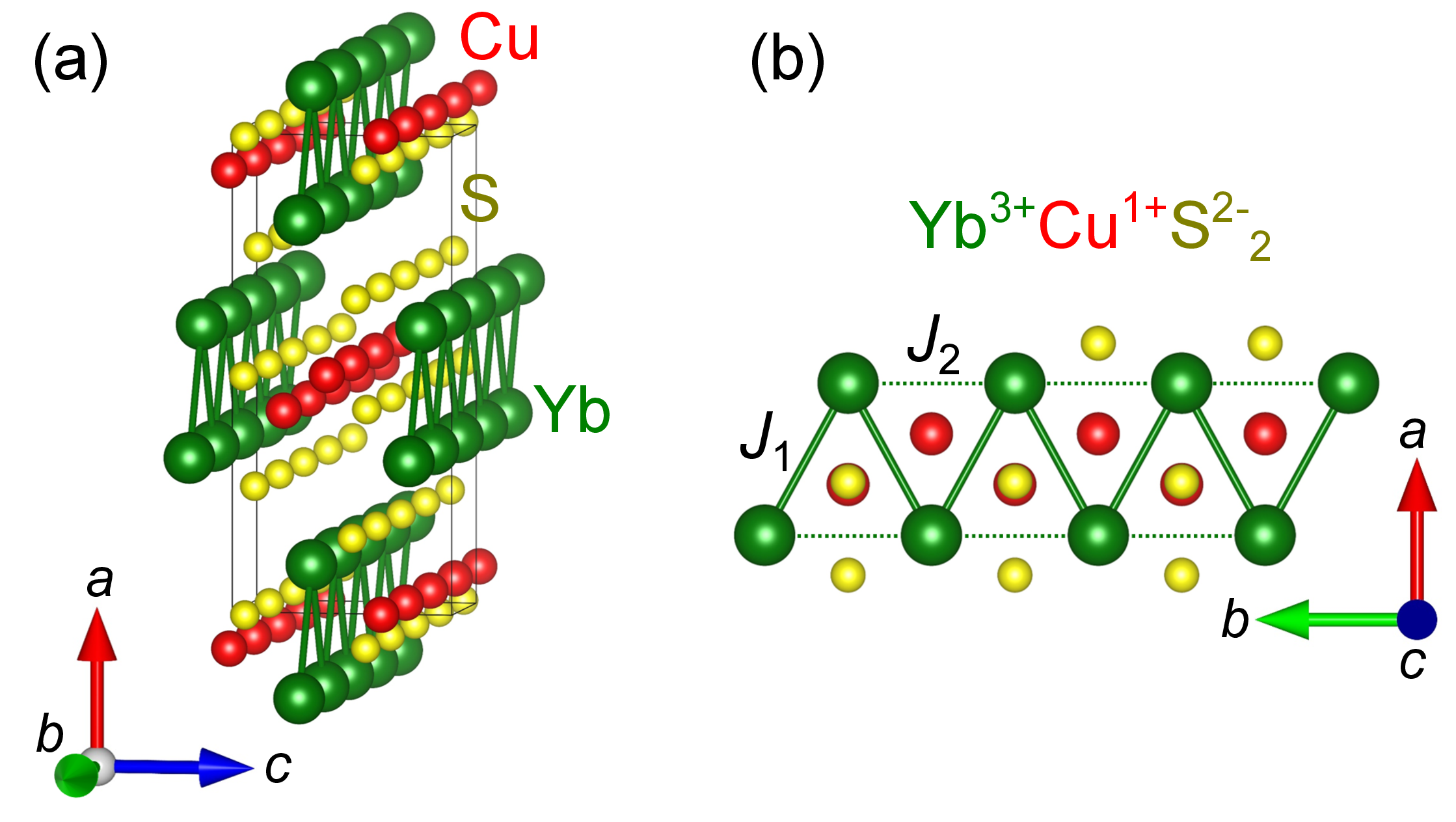} 
\caption{(a) Crystal structure of YbCuS$_2$ with the space group $Pnma$ (No. 62, $D_{2h}^{16}$). The black box represents the unit cell. 
(b) View from the $c$-axis; zigzag chains along the $b$-axis are formed by Yb atoms. The structural image was produced using the VESTA program~\cite{vesta}.}
\label{structure}
\end{figure}

4$f$-electron-based magnetic semiconductor YbCuS$_2$ has recently attracted much attention as one of the frustrated magnetic semiconductor with anisotropic exchange interactions~\cite{Ohmagari1, Ohmagari2, Hori2022, Hori2023, Saito1, Saito2, Saito3}.
This system has the Yb$^{3+}$ zigzag chains as shown in Fig.~\ref{structure}, where frustration effect driven by the competition between the nearest-neighbor exchange interaction $J_1$ and the next-nearest-neighbor exchange interaction $J_2$ is expected.
Indeed, several unique properties resulting from this frustration effect have been consistently observed~\cite{Ohmagari1, Ohmagari2, Hori2022, Hori2023}.
This compound shows a first-order magnetic transition at $T_{\rm N} \sim 0.95$~K characterized by the strong divergence of the specific heat and coexistence of paramagnetic (PM)  and antiferromagnetic (AFM) NQR signals at around $T_{\rm N}$.
Below $T_{\rm N}$, an incommensurate AFM structure is suggested by the $^{63/65}$Cu-NQR measurement~\cite{Hori2023} and confirmed by the neutron scattering experiment~\cite{Onimaru_YbCuS2_neutron}.
Moreover, the ordered moment is much smaller than that anticipated from the crystalline electric field ground states.
The $^{63}$Cu-NQR nuclear spin-lattice relaxation rate $1/T_1$ exhibits a broad maximum at $T_{\rm max} \sim 50$~K, indicating the presence of a short-range order.

The most striking feature is the $T$-linear behavior of $1/T_1$ below 0.5~K.
This $T$-linear behavior cannot be explained by conventional magnon excitation, and indicates the gapless excitations by charge-neutral quasiparticles in YbCuS$_2$~\cite{Hori2023}.
However, the origin and characteristics of the gapless excitations remain unclear.
After the experimental observation, it has been theoretically proposed that quantum anisotropic exchange interactions $\Gamma$-term originating from the Yb zigzag chains may give rise to gapless emergent quasiparticles called ``nematic particles", which explain the NQR results~\cite{Saito1, Saito2, Saito3}.
In this proposed spin model, the Hamiltonian is given as
\begin{flalign}
\mathcal{H}=\sum_j \sum_{\eta=1,2} J_\eta \boldsymbol{S}_j \cdot \boldsymbol{S}_{j+\eta}+\Gamma_\eta\left(S_j^x S_{j+\eta}^y+S_j^y S_{j+\eta}^x\right),
\label{eq:zigzag_gamma}
\end{flalign}
where 
$J_{\eta}$ and $\Gamma_\eta$ are the Heisenberg and anisotropic exchange interactions between nearest ($\eta = 1$) and next-nearest ($\eta = 2$) spins.
The $x$, $y$, and $z$ axes correspond to the $a$, $c$, and $b$ axes in YbCuS$_2$, respectively.
This model predicts a ground-state phase diagram featuring a \textit{quantum Lifshitz multicritical point}, where five distinct phases converge at a point; a uniform ferromagnet, two antiferromagnets, Tomonaga-Luttinger liquid, and a dimer-singlet coexisting with nematic order~\cite{Saito2, Saito3}.
Therefore, experimental investigations on YbCuS$_2$ by changing interaction parameters $J_{\eta}$ and $\Gamma_\eta$ are promising for discovering intriguing phenomena.

In this paper, we perform $^{63/65}$Cu-NQR measurements by applying hydrostatic pressures, since it is a cleanest tuning method in the system.
The transition temperature $T_{\rm N}$
increases from 0.95 K at ambient pressure to 1.17 K at $P =1.6$~GPa
by applying pressures.
In addition, the magnetic structure below $T_{\rm N}$ changes from an incommensurate helical order in the $ac$ plane at ambient pressure to a commensurate $c$-axis antiferromagnetic order under high pressure.
This high-pressure magnetic structure can be regarded as an odd-parity magnetic multipole order.
The value of $1/T_1T$, which indicates gapless quasiparticle excitations, decreases by applying pressure.
These results indicate that pressure evolution in the magnetic interactions between the Yb ions changes the magnetic structure and suppresses the gapless quasiparticle excitations,
which are consistently understood by the recent theoretical model~\cite{Saito1, Saito2, Saito3} with the decrease in $J_2/J_1$.

\section{Experimental Results}

\begin{figure}[t!]
\centering
\includegraphics[width=8cm]{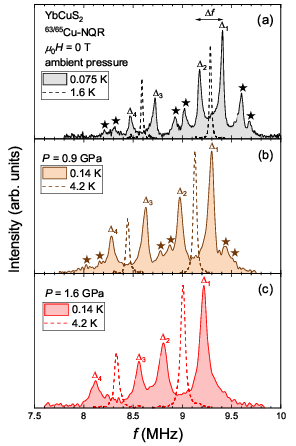} 
\caption{The $^{63/65}$Cu-NQR spectra of YbCuS$_2$ at (a) ambient pressure, (b) 0.9~GPa, and (c) 1.6~GPa; dashed and solid curves denote the spectra above and below $T_{\rm N}$, respectively.
}
\label{spectrum}
\end{figure}

The $^{63/65}$Cu-NQR spectra measured above (below) $T_{\rm N}$ under various pressures are presented with dashed (solid) curves in Fig.~\ref{spectrum}.
Above $T_{\rm N}$, a single sharp peak was observed.
As pressure increases, the peak frequency decreases.
This indicates that the electric field gradient (EFG) around the observed Cu nuclei varied owing to compression.
In nuclear magnetic resonance (NMR) and NQR measurements, there are two EFG parameters; the quadrupole frequency along the principal axis of the EFG $\nu_{zz}$ and the asymmetry parameter $\eta$.
From the NMR spectrum above $T_{\rm N}$ at 1.6 GPa, we confirmed that $\eta$ remains unchanged, while $\nu_{zz}$ decreases under pressure.
In addition, the line width of the NQR spectrum slightly increases with applying pressure, indicating the inhomogeneity of pressure.
Below $T_{\rm N}$, the NQR signal splits, reflecting the appearance of the internal magnetic fields $H_{\rm int}$.
As previously reported~\cite{Hori2023}, 10 peaks were observed at ambient pressure below $T_{\rm N}$.
However, as pressure increases, the relative intensity of the star-marked peaks against the main peaks ($\bigtriangleup_1$-$\bigtriangleup_4$) decreases.
At 1.6~GPa, the star-marked peaks were undetectable, and only 4 peaks remain.
These results indicate that the magnetic structure changed under pressure.
Possible magnetic structures will be discussed later.

\begin{figure}[t!]
\centering
\includegraphics[width=9cm]{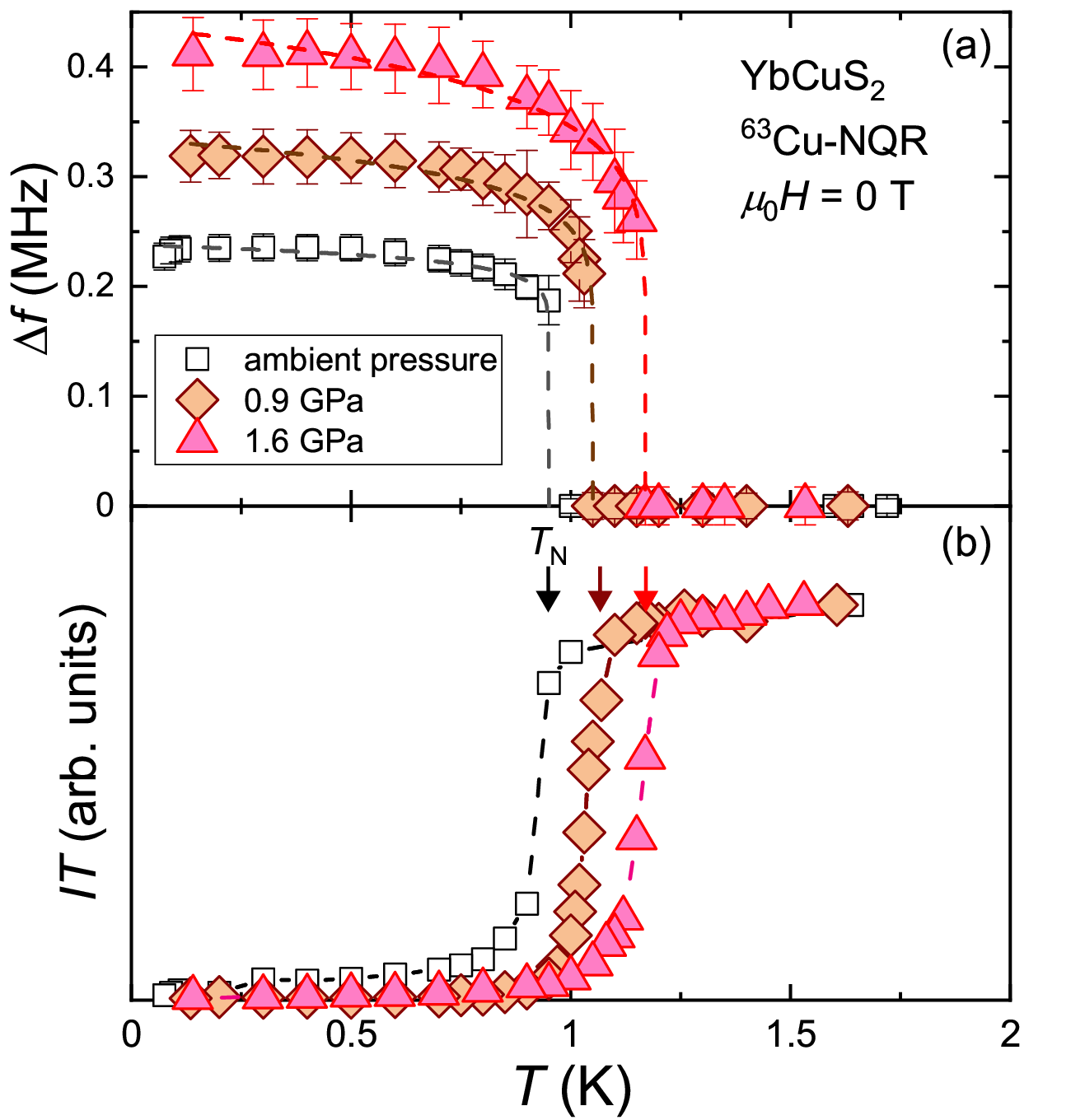} 
\caption{(a) Temperature dependence of frequency difference of splitting peaks $\Delta f$ on YbCuS$_2$ (b) Temperature dependence of the products of the $^{63}$Cu-NQR peak intensity and the temperature $I(T) T$ on YbCuS$_2$; the squares, diamonds and triangles denote $I(T) T$ or $\Delta f$ at ambient pressure, 0.9~GPa and 1.6~GPa, respectively.
}
\label{IT}
\end{figure}

To track the pressure dependence of the transition temperatures $T_{\rm N}$, we investigated the temperature dependence of frequency difference of splitting peaks $\Delta f$ and
the NQR peak intensity $I (T)$ as shown in Fig.~\ref{IT}.
Here, $I(T)$ was measured at frequencies of NQR peak above $T_{\rm N}$, and $\Delta f$ represents the frequency difference between the two peaks denoted by $\bigtriangleup_1$ and $\bigtriangleup_2$ in Fig.~\ref{spectrum}.
In the PM state, the product of the NQR peak intensity and the temperature $I(T) T$ is almost constant against temperature.
However, deviations from $I(T) T = \text{const.}$ are observed when there is splitting or broadening of the spectrum below $T_{\rm N}$.
We define $T_{\rm N}$ as the temperature below which $\Delta f$ increases and $I(T) T$ decreases rapidly. 
As previously reported~\cite{Hori2023}, it was confirmed that $T_{\rm N} \sim 0.95$~K at ambient pressure. 
By applying pressure, $T_{\rm N}$ increases to 1.07~K at 0.9~GPa and 1.17~K at 1.6~GPa.

In the previous NQR study at ambient pressure~\cite{Hori2023}, we observed the sudden splitting of signals and coexistence of the PM and AFM signals at around $T_{\rm N}$.
We assume that $\Delta f$ is proportional to the internal magnetic field $H_{\rm int}$  and follows $\Delta f(T) = \Delta f(0)[(T_{\rm N} - T)/T_{\rm N}]^{\beta}$ as shown in the dashed curve of Fig~\ref{IT}(a).
The estimated critical exponent $\beta$ is 0.05, which is substantially smaller than the conventional mean-field value 0.5.
These results indicate that the AFM transition is an first-order phase transition at ambient pressure~\cite{Hori2023}.
Here, we investigated how the first-order feature of phase transition in YbCuS$_2$ changes under pressure.
As pressure increases, $\beta$ increases to 0.09 at 0.9~GPa and 0.12 at 1.6~GPa, which is still much smaller than 0.5.
Note that the PM and AFM signals still coexist in a certain temperature region under pressure.
These results also suggests that the first-order nature is robust under pressure.

\begin{figure*}[t]
\centering
\includegraphics[width=11cm]{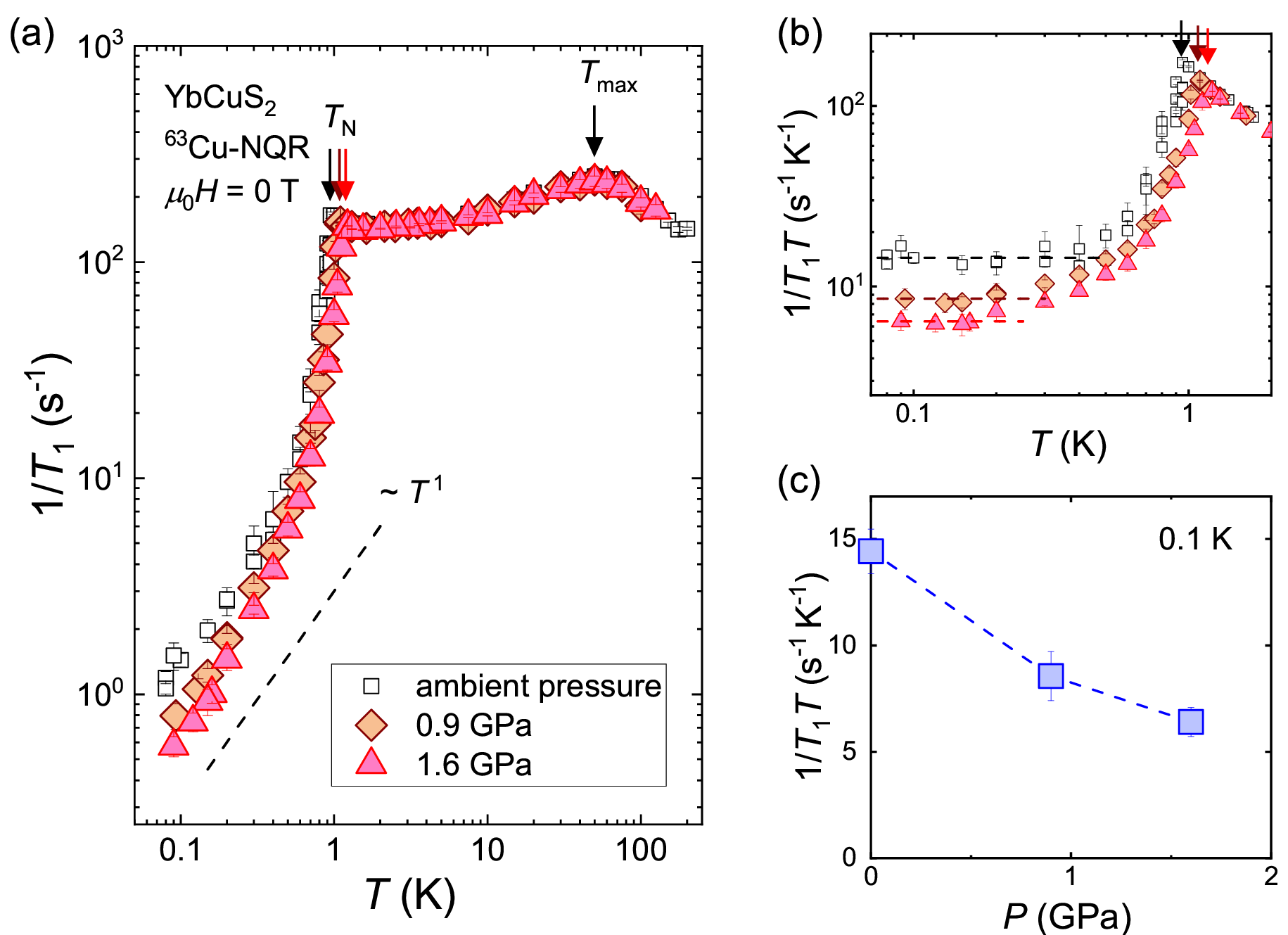} 
\caption{
(a) Temperature dependence of $1/T_1$ on YbCuS$_2$; the squares, diamonds and triangles denote $1/T_1$ at ambient pressure, 0.9~GPa and 1.6~GPa, respectively.
(b) Temperature dependence of $1/T_1T$ on YbCuS$_2$.
(c) Pressure dependence of $1/T_1T$ at 0.1~K on YbCuS$_2$.
}
\label{1_T1}
\end{figure*}

Figure~\ref{1_T1}(a) shows the temperature dependence of $1/T_1$ measured by the $^{63}$Cu-NQR in YbCuS$_2$ under various pressure values.
Despite applying pressure, no significant change in the anomaly at $T_{\rm max} \sim 50$~K was observed. 
This indicates that the anomaly at $T_{\rm max}$ has robust characteristic against the pressure.
The increase in $T_{\rm N}$ due to pressure are also confirmed from the temperature dependence of $1/T_1$.
Even with pressure applied, a gapless behavior with $1/T_1 \sim T^{1}$ is observed far below $T_{\rm N}$,
although its value decreases from $1/T_1T \sim 14$~s$^{-1}$K$^{-1}$ at ambient pressure to 6~s$^{-1}$K$^{-1}$ at 1.6~GPa, as shown in Fig~\ref{1_T1}(b) and \ref{1_T1}(c).

\section{Discussion}

\begin{figure*}[t!]
\centering
\includegraphics[width=18cm]{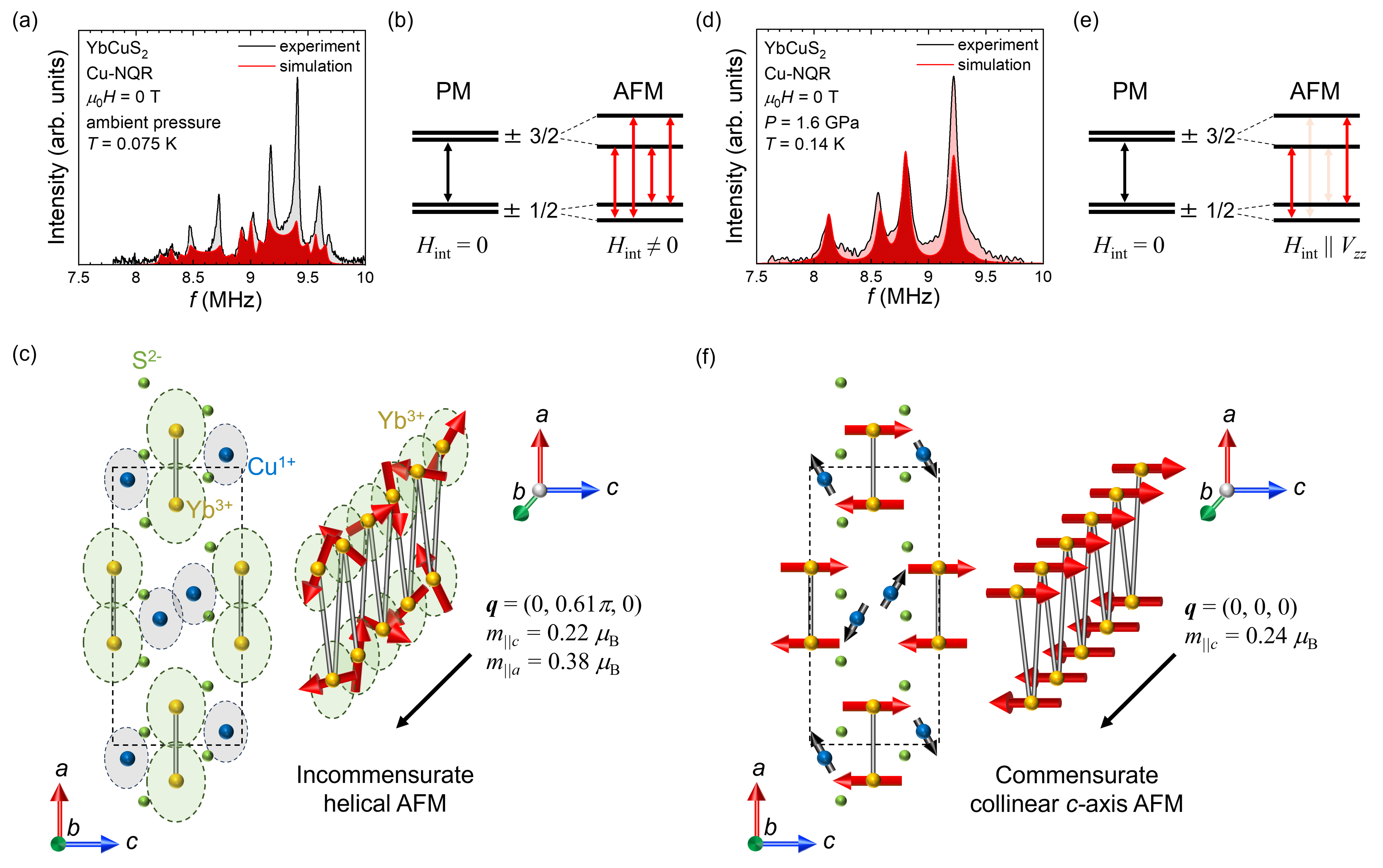} 
\caption{(a) The $^{63/65}$Cu-NQR spectra of YbCuS$_2$ at ambient pressure; the solid black and red curve are the experimental data and simulation of the NQR spectra, respectively.
(b) The energy-level diagram for the Cu nuclear spin at ambient pressure.
(c) The possible magnetic structure explaining the NQR spectrum at ambient pressure;
the green and gray ellipses represent the incommensurate ordered moments and the distribution of the internal magnetic fields, respectively.
(d) The $^{63/65}$Cu-NQR spectra of YbCuS$_2$ at 1.6~GPa.
(e) The energy-level diagram for the Cu nuclear spin at 1.6~GPa.
(f) The possible magnetic structure explaining the NQR spectrum at 1.6~GPa; the red and black arrows show the Yb$^{3+}$ ordered moments and the internal magnetic fields at Cu nuclei, respectively. 
}
\label{simulation}
\end{figure*}

Here, we discuss the possible magnetic structure in the AFM state of YbCuS$_2$ at 1.6~GPa on the basis of the simulation of the $^{63/65}$Cu-NQR spectra below $T_{\rm N}$ shown in Fig.~\ref{simulation}.
We considered classical dipolar fields and the principal axis of the EFG $V_{zz}$, evaluated by WIEN2K calculation~\cite{WIEN2k} with the density functional theory as reported in previous study~\cite{Hori2023}.
At ambient pressure, $^{63/65}$Cu-NQR of YbCuS$_2$ splits into 10 peaks due to the multiple $H_{\rm int}$ and wide distribution of  $H_{\rm int}$ at the Cu sites,  which is induced by an incommensurate magnetic structure as reported previously~\cite{Hori2023}.
Recently, the elastic neutron scattering measurement has been performed~\cite{Onimaru_YbCuS2_neutron}, revealing
helical magnetic order with a incommensurate wave vector $\bm{q} = (0, 0.61 \pi, 0)$ along the $b$ axis and ordered moments rotating on the elliptical $ac$ plane.
The size of the ordered moment is $m_{\parallel c}=0.22$~$\mu_{\rm B}$ along the $c$ axis and $m_{\parallel a}=0.38$~$\mu_{\rm B}$ along the $a$ axis as shown Fig.~\ref{simulation}(c).
This magnetic structure is consistent with the $^{63/65}$Cu-NQR spectrum at ambient pressure as shown in Fig.~\ref{simulation}(a), although the intensity of all peaks cannot be fully reproduced.

On the other hand, the 4 peaks observed at 1.6~GPa suggest a single $H_{\rm int}$ parallel to $V_{zz}$ at the Cu nuclei, which arises from certain resonance transitions, as shown in Fig.~\ref{simulation}(e).
Therefore, it is reasonable to consider that the incommensurate magnetic order changes to a commensurate one under pressure.
We performed spectrum simulations on several magnetic structures and found that a commensurate collinear AFM state with $\bm{q} = (0, 0, 0)$ and ordered moments $m_{\parallel c} = 0.24$~$\mu_{\rm B}$ per Yb$^{3+}$ yields $H_{\rm int}$ parallel to $V_{zz}$ as shown in Fig.~\ref{simulation}(f).
In the mean-field model of the quasi-one-dimensional chains, $T_{\rm N}$ and the moment size $m$ are related to the inter-chain interaction $J_{3}$ as follows $T_{\rm N} \sim J_{3}$ and $m \sim \sqrt{J_{3}}$~\cite{1D_Neel_moment_1, 1D_Neel_moment_2, 1D_Neel_moment_3}.
Based on this formula, the observed pressure change in $T_{\rm N} $ of 1.17~K/0.95~K $\sim 1.2$ leads to a change in $m$ of 1.1 times.
It is consistent with the observed variation of $m_{\parallel c}$ ($0.24\mu_{\rm B}/0.22\mu_{\rm B} = 1.1$).

Interestingly, the magnetic structure at 1.6~GPa can be regarded as an odd-parity magnetic multipole order~\cite{Hayami, Yatsushiro, Hayami_JPSJ_review}, where both space-inversion ($\mathcal{P}$) and time-reversal ($\mathcal{T}$) symmetries are violated but space-time inversion ($\mathcal{PT}$) symmetry is preserved below $T_{\rm N}$, although both $\mathcal{P}$ and $\mathcal{T}$ symmetries are preserved above $T_{\rm N}$.
In such a case, magnetic toroidal dipoles and magnetic quadrupole are expected to be active~\cite{Hayami, Yatsushiro, Wanatabe_Yanase_1, Wanatabe_Yanase_2, Hayami_JPSJ_review} and intriguing functionalities such as cross-correlation responses including magnetoelectric effects and non-reciprocal thermal conductivity may arise, which have been recently studied from experimental and theoretical sides~\cite{Hayami, Yatsushiro, Wanatabe_Yanase_1, Wanatabe_Yanase_2, Hayami_JPSJ_review,  Saito_UNi4B, ME_Ce3TiBi5, Nonlinear_Hall_CuMnAs}.
Therefore, YbCuS$_2$ provides a promising platform to study an odd-parity multipole physics and cross-correlation responses.

\begin{figure}[t]
\centering
\includegraphics[width=9cm]{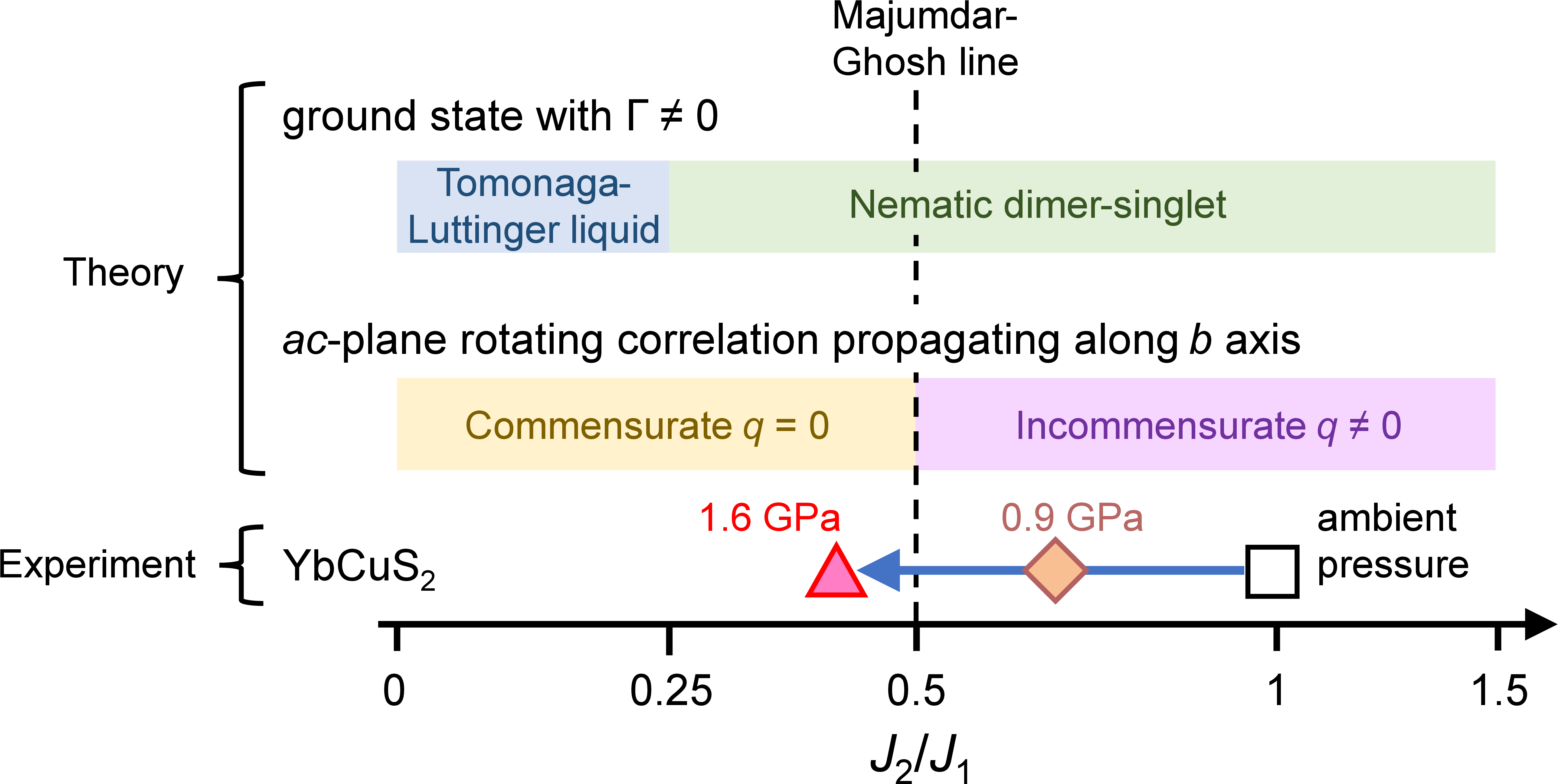} 
\caption{
Schematic image of the pressure change in the magnetic correlation on YbCuS$_2$;
the theoretical basis is provided in Refs.~\cite{Saito2} and \cite{Saito3}.
The ground state presented here represents the states at $\Gamma_
1/J_1 = \Gamma_2/J_2 = 0.01$ in the Eq.~(\ref{eq:zigzag_gamma}), where a dimer-singlet state coexisting with nematic order (nematic dimer-singlet) transitions to a Tomonaga-Luttinger liquid state at $J_2/J_1 \sim 0.25$.
The detailed ground-state phase diagram for this model is also shown in Fig.~1 of Refs.~\cite{Saito2} and \cite{Saito3}.
The $ac$-plane rotating incommensurate correlation propagating along the $b$ axis changes to a commensurate one at $J_2/J_1 = 0.5$ (Majumdar-Ghosh line).
Our experimental results for YbCuS$_2$ suggest that applying pressure reduces $J_2/J_1$ from 1 at ambient pressure to less than 0.5 at 1.6~GPa.
}
\label{phase_diagram}
\end{figure}

We discuss the origin of the changes in magnetic structure and gapless excitation under pressure.
In the above-mentioned theoretical zigzag chain model including small anisotropic interaction $\Gamma$,
long-range magnetic order cannot be realized in the region with $J_2/J_1 \sim 1$ relevant to YbCuS$_2$~\cite{Saito1, Saito2, Saito3}.
Therefore, it is natural to attribute the discrepancy between the theoretical model and the experimental ground state to the inter-chain coupling $J_3$, which stabilizes a magnetic order.
However, the realized magnetic structure is closely related to the intra-chain interactions. 
Indeed, the model with $J_2/J_1 \sim 1$ and small $\Gamma$ predict the $ac$-plane rotating incommensurate correlation propagating along the $b$ axis, 
which is relevant to the experimental results of the helical magnetic structure at ambient pressure~\cite{Saito3}.
In addition, it has been suggested that reducing the ratio $J_2/J_1$ to less than 0.5 results in the transition from an incommensurate magnetic correlation to a commensurate one~\cite{Saito3}.
This commensurate correlation seems to be consistent with the experimental result of the commensurate collinear antiferromagnetic state with ordered moments along the $c$ axis under pressure.
Thus, based on the theoretical anisotropic zigzag chain model, we propose that applying pressure reduces the ratio $J_2/J_1$ to
less than 0.5, as illustrated in Fig.~\ref{phase_diagram}.
As the changes in the $J_1$ and $J_2$ parameters can be related to crystal structure changes, the crystal structural analysis under pressure is essential to support our proposal.

The value of $1/T_1T$ at low temperatures is suppressed by applying pressure.
Here, we propose two possible origins about the gapless excitation.
First, the intra-chain interaction parameters may affect the gapless excitations as well as low-temperature magnetic structure.
In the zigzag chain models with $\Gamma$, there exists an exact solution of the gapped singlet product state at $J_2/J_1 = 0.5$ (Majumdar-Ghosh line) as a singularity line, where nematic orders and nematic particles disappear~\cite{Saito2}.
This may prevent the appearance of gapless quasiparticle excitation.
It is experimental challenge whether such a line is detectable by actual measurements.
Second, the increase in $T_{\rm N}$ due to pressure might reduce the fluctuations, leading to the suppression of gapless excitations.
The increase in $T_{\rm N}$ suggests that $J_3$ becomes larger by applying pressure.
In typical quantum critical phenomena~\cite{QCP_Sachdev, QCP_Coleman, QCP_heavy-fermion}, increasing $T_{\rm N}$ to stabilize the ordered state moves the system further from the quantum critical point, thereby suppressing magnetic fluctuations.
In this case, applying additional pressure to increase $T_{\rm N}$ would be expected to further suppress the fluctuations inducing gapless excitations.
Note that, in either scenario, our results 
strongly support
that the origin of the gapless excitations is related to the intra-chain interactions, which contribute to quantum fluctuations.
The findings from our systematic study emphasize the significance of the one-dimensional nature of these systems.

\section{Conclusion} 

In conclusion, we report $^{63/65}$Cu-NQR measurements under hydrostatic pressures.
The transition temperature $T_{\rm N}$ increases from 0.95 K to 1.17 K under pressures up to 1.6 GPa.
In addition, the magnetic structure below $T_{\rm N}$ changes from an incommensurate helical order along the $b$ axis with the ordered moment in the $ac$ plane at ambient pressure to a commensurate $c$-axis antiferromagnetic order under pressure.
This commensurate magnetic structure can be regarded as an odd-parity magnetic multipole order.
The value of $1/T_1T$, which indicates gapless quasiparticle excitations, decreases by applying pressure.
These results suggest that pressure evolution in the magnetic interactions between Yb ions changes the magnetic structure and suppresses the gapless quasiparticle excitations.
Interestingly, the present findings from our systematic study seems to be consistent with the recent theoretical models of the one-dimensional zigzag chain with anisotropic exchange interactions $\Gamma$ term and highlights the richness of the one-dimensional nature of this system.

\section{Method}

\subsection{Crystal growth and hydrostatic pressure}
Polycrystalline samples of YbCuS$_{2}$ were synthesized by the melt-growth method.
The hydrostatic pressure is applied using a piston cylinder–type cell made of NiCrAl and CuBe alloys.
Daphne 7373 was used as a pressure medium.
The applied pressure $P$ was estimated by the superconductiving transition temperature $T_{\rm c}(P)$ of Pb with the formula of $P=[T_{\rm c}(0)-T_{\rm c}(P)]/0.364$~\cite{pressure_Pb}.

\subsection{NQR measurements}
A conventional spin-echo technique was used for the  $^{63/65}$Cu-NQR measurements.
A $^3$He-$^4$He dilution refrigerator was used for the  $^{63/65}$Cu-NQR measurements under pressure down to 0.1 K.
$^{63}$Cu-$1/T_1$ was evaluated by fitting the relaxation curve of the nuclear magnetization after the saturation to a theoretical function for the nuclear spin $I = 3/2$.
$1/T_1$ was determined by a single relaxation component down to $T_{\rm N}$. 
However, short component appears in the relaxation curve below $T_\mathrm{N}$.
Thus, we picked up the slowest components as described in previous study~\cite{Hori2023}.

\section{Data availability}
The data that support the findings of this study are available from the corresponding author upon reasonable request.

\section{Author contributions}
F.H. and K.I. designed the research.
F.H., H.M., S.K., and K.I. performed NQR measurements.
H.S. and T.O. synthesized and characterized the bulk samples.

\section{Competing interests}
The authors declare no competing interests.

\begin{acknowledgments}
The authors would like to thank H. Saito, and C. Hotta for valuable discussions. 
This work was supported by Grants-in-Aid for Scientific Research (KAKENHI Grant Nos. JP20KK0061, JP20H00130, JP21K18600, JP22H04933, JP22H01168, JP23H01124, JP23K22439, JP23K25821, JP23H04866, JP23H04870, JP23KJ1247 and JP24K00574) from the Japan Society for the Promotion of Science, by JST SPRING (Grant No. JPMJSP2110) from Japan Science and Technology Agency, by research support funding from the Kyoto University Foundation, by ISHIZUE 2024 of Kyoto University Research Development Program, and by Murata Science and Education Foundation.
In addition, liquid helium is supplied by the Low Temperature and Materials Sciences Division, Agency for Health, Safety and Environment, Kyoto University.
\end{acknowledgments}

\clearpage

\bibliographystyle{apsrev4-2}

\end{document}